\documentclass[conference,a4paper]{IEEEtran}

\usepackage[utf8]{inputenc}
\usepackage[T1]{fontenc}
\usepackage{cite}
\usepackage{amsmath,amssymb,amsfonts}
\usepackage{graphicx}
\usepackage{textcomp}
\usepackage{xcolor}
\usepackage{booktabs}
\usepackage{multirow}
\usepackage{url}
\usepackage[hypertexnames=false]{hyperref}
\usepackage{algorithm}
\usepackage{algorithmic}
\usepackage{float}
\usepackage{tikz}
\usetikzlibrary{shapes.geometric, arrows.meta, positioning, fit, backgrounds, patterns}
\usepackage{pgfplots}
\pgfplotsset{compat=1.18}
\definecolor{tpblue}{HTML}{2A78D6}
\definecolor{fporange}{HTML}{EB6834}
\definecolor{condteal}{HTML}{4C9A8E}
\definecolor{indgray}{HTML}{D9D8D4}

\def\BibTeX{{\rm B\kern-.05em{\sc i\kern-.025em b}\kern-.08em
    T\kern-.1667em\lower.7ex\hbox{E}\kern-.125emX}}

\begin{document}

\title{Validating AI-Discovered Cloud Threats on Live Infrastructure: A Multi-Model, Multi-Cloud Study}

\author{\IEEEauthorblockN{Nicholas Pecka\IEEEauthorrefmark{1}\IEEEauthorrefmark{2}, Lotfi Ben Othmane\IEEEauthorrefmark{1}, and Ren\'ee Bryce\IEEEauthorrefmark{1}}
\IEEEauthorblockA{\IEEEauthorrefmark{1}University of North Texas, Denton, TX, USA\qquad
\IEEEauthorrefmark{2}Red Hat, Raleigh, NC, USA}
}

\maketitle

\begin{abstract}
Cloud infrastructure threat modeling lacks empirical validation of AI-discovered threats. We present a three-phase methodology of multi-agent threat discovery, automated exploit synthesis, and non-destructive live validation, and apply it to three commercial foundation models on two cloud architectures, validating each threat natively on the cloud (AWS, GCP, or Azure) it was discovered against. Of 233 AI-discovered threats, 103 were confirmed and 34 refuted (75.2\% precision, 95\% CI [67.3, 81.7]); 44 were conditional (the named mechanism refuted but an equivalent impact shown through another path) and 52 indeterminate. Multi-model discovery helped through breadth, not agreement: the three models together found 2.2$\times$ the confirmed threats of the best single model at comparable precision, while threats reported by several models were confirmed no more often than single-model ones. False positives were not architecture-specific: they asserted access or system state the deployment did not have. Threats coded in advance as observable configuration resolved cleanly more often than multi-step behavioral threats (72.5\% vs.\ 51.6\%), which were more often conditional, suggesting that models more often misidentify the exploitation mechanism than the underlying weakness. AI-assisted discovery is a practical input to security workflows when paired with empirical validation.
\end{abstract}


\section{Introduction}

Cloud-native systems introduce infrastructure-level threats that traditional tools miss. Architectural composition creates emergent vulnerabilities: trust-boundary violations through credential sharing, privilege-escalation chains across cloud services, and supply-chain attacks on continuous integration and continuous deployment (CI/CD) pipelines~\cite{pan2024cicd}. Manual threat modeling does not scale, and automated scanners analyze components in isolation without cross-resource attack paths~\cite{pecka2026semiautomated}; infrastructure-as-code (IaC) scripts frequently carry misconfigurations~\cite{rahman2019sevensins}, yet static analysis yields high false-positive (FP) rates~\cite{wang2024falsep}. Critical architectural threats thus stay undetected until exploited in production.

Foundation models offer semantic reasoning over cloud configurations, jointly analyzing IaC, network topology, and access policies; large language models (LLMs) have been applied across security tasks including vulnerability detection~\cite{yang2025llmsec}. But threats generated by artificial intelligence (AI) may be false positives from incorrect assumptions about system state or hallucinated attack paths~\cite{zhang2023hallucination}, so production deployment requires empirical validation to quantify accuracy and identify systematic failure modes.

Despite growing adoption of AI security tools, the research community lacks empirical validation of AI-discovered cloud threats. Existing work on LLM-assisted security focuses on penetration testing~\cite{deng2023pentestgpt} and threat model generation~\cite{Crossman2025Auspex,adams2023stridegpt}, but does not validate whether discovered threats are actually exploitable against deployed infrastructure. Critical open questions include: What is the true positive (TP) rate of AI-discovered threats? How does multi-model consensus affect accuracy? Do severity ratings predict exploitability? Does accuracy depend on the architecture? Without answers, organizations cannot confidently integrate AI threat discovery into security workflows.

This paper addresses these questions by testing, through live exploitation, 233 threats discovered by three commercial foundation models across three cloud platforms (Amazon Web Services (AWS), Google Cloud Platform (GCP), and Microsoft Azure) and two representative architectures.

Contributions:
\begin{itemize}
\item A three-phase methodology for empirically testing AI-discovered cloud threats through automated exploit synthesis and non-destructive live validation against production-like infrastructure, with layered verification: live-evidence verdicts, an independent model re-review, and expert adjudication.
\item Native live validation of 233 AI-discovered threats, each on the cloud it was discovered against (six deployments across AWS, GCP, and Azure), showing 75.2\% precision among resolved threats.
\item Evidence that multi-model discovery improves results through coverage rather than agreement: three models together found 2.2$\times$ the confirmed threats of the best single model at comparable precision, while cross-model agreement did not predict correctness.
\item A characterization of AI threat-detection error: false positives asserted access or system state the deployment did not have rather than following an architecture, and a blind a-priori coding suggests multi-step behavioral threats are the ones whose mechanism is most often wrong.
\item Practical guidance for deploying AI threat detection, covering model selection, observability-based triage, and the limits of model consensus.
\end{itemize}

\section{Related Work}

Recent tools use LLMs for threat modeling: Auspex~\cite{Crossman2025Auspex} generates STRIDE matrices from architecture diagrams, STRIDE GPT~\cite{adams2023stridegpt} produces STRIDE models, attack trees, and DREAD scores from a textual description, and ThreatModeling-LLM~\cite{wu2024threatmodelingllm} fine-tunes LLMs on a Microsoft Threat Modeling Tool benchmark for banking systems. All evaluate LLM output against descriptions, diagrams, or expert benchmarks; none test whether the generated threats are exploitable in a running system. Related efforts apply LLMs to penetration testing~\cite{deng2023pentestgpt} and vulnerability assessment~\cite{yang2025llmsec,zhang2025llmvuln}, with one project-scale study reporting high false-discovery rates on real code~\cite{zhang2025llmvuln}. Our work differs in three ways: threats are discovered from configuration extracted from deployed cloud infrastructure rather than from a description, every threat is tested against that live infrastructure, and multiple models are compared on the same systems across three clouds.

Production AI-security systems show the stakes. Traust~\cite{traust2024}, a workflow engine for organization-wide audits, adopts an ``asymmetric-caution doctrine'' in which machines may confirm findings but only humans may dismiss them, documenting the error rates behind that choice; multi-agent security systems also face malicious-agent and inter-agent-trust challenges~\cite{zhang2025multiagent}. This false-positive challenge motivates our test of whether multi-model consensus improves accuracy.

Combining several models by voting is an established way to improve security detection, e.g., ensemble intrusion detection with adaptive voting~\cite{gao2019adaptiveensemble}. For LLMs, agreement has been used to signal correctness: majority voting over sampled reasoning paths improves accuracy~\cite{wang2023selfconsistency}, inconsistency flags hallucinated facts~\cite{manakul2023selfcheckgpt}, and multi-instance debate reduces factual errors~\cite{du2023debate}. Since LLMs produce plausible but incorrect output~\cite{zhang2023hallucination}, such signals are attractive for threat discovery; yet agreement \emph{across different foundation models} has not been tested as a signal of whether discovered threats are real. We test it and find it weaker than these results suggest.

The taxonomy used here~\cite{pecka2026semiautomated} includes ML-specific threats spanning model theft~\cite{tramer2016stealing}, training poisoning~\cite{gu2017badnets}, and inference attacks~\cite{shokri2017membership}; asset-centric AI threat modeling has also been proposed~\cite{kumar2024stride}. That work targets algorithm-level robustness, whereas we address infrastructure-level ML threats from deployment misconfigurations.

\section{Methodology}
\label{sec:method}

\begin{figure}[t]
\centering
\resizebox{\columnwidth}{!}{%
\begin{tikzpicture}[
    node distance=0.6cm and 1cm,
    box/.style={rectangle, draw, minimum width=2cm, minimum height=0.6cm, align=center},
    phase/.style={rectangle, draw, thick, minimum width=3cm, minimum height=1.5cm, align=center},
    arrow/.style={->, >=stealth, thick}
]
\node[box, fill=orange!20, minimum width=3.5cm] (cases) at (0,0) {Case Studies\\CS1 \& CS2};
\node[box, fill=green!15, below=0.5cm of cases, minimum width=3.5cm] (platforms) {Platforms\\AWS / GCP / Azure};
\node[box, fill=blue!15, below=0.5cm of platforms, minimum width=3.5cm] (models) {Models\\Gemini / Claude / GPT};
\node[phase, fill=blue!5, below=0.8cm of models] (p1) {Phase 1\\Discovery\\(3 agents)};
\node[phase, fill=yellow!10, right=1.2cm of p1] (p2) {Phase 2\\Synthesis\\(exploits)};
\node[phase, fill=green!5, right=1.2cm of p2] (p3) {Phase 3\\Native\\validation};
\draw[arrow] (cases) -- (platforms);
\draw[arrow] (platforms) -- (models);
\draw[arrow] (models) -- (p1);
\draw[arrow] (p1) -- (p2);
\draw[arrow] (p2) -- (p3);
\end{tikzpicture}}
\caption{Three-phase methodology: two case studies on three clouds; multi-agent discovery by three foundation models; then automated exploit synthesis and native live validation classifying each threat as TP, FP, conditional, or indeterminate.}
\label{fig:methodology}
\end{figure}
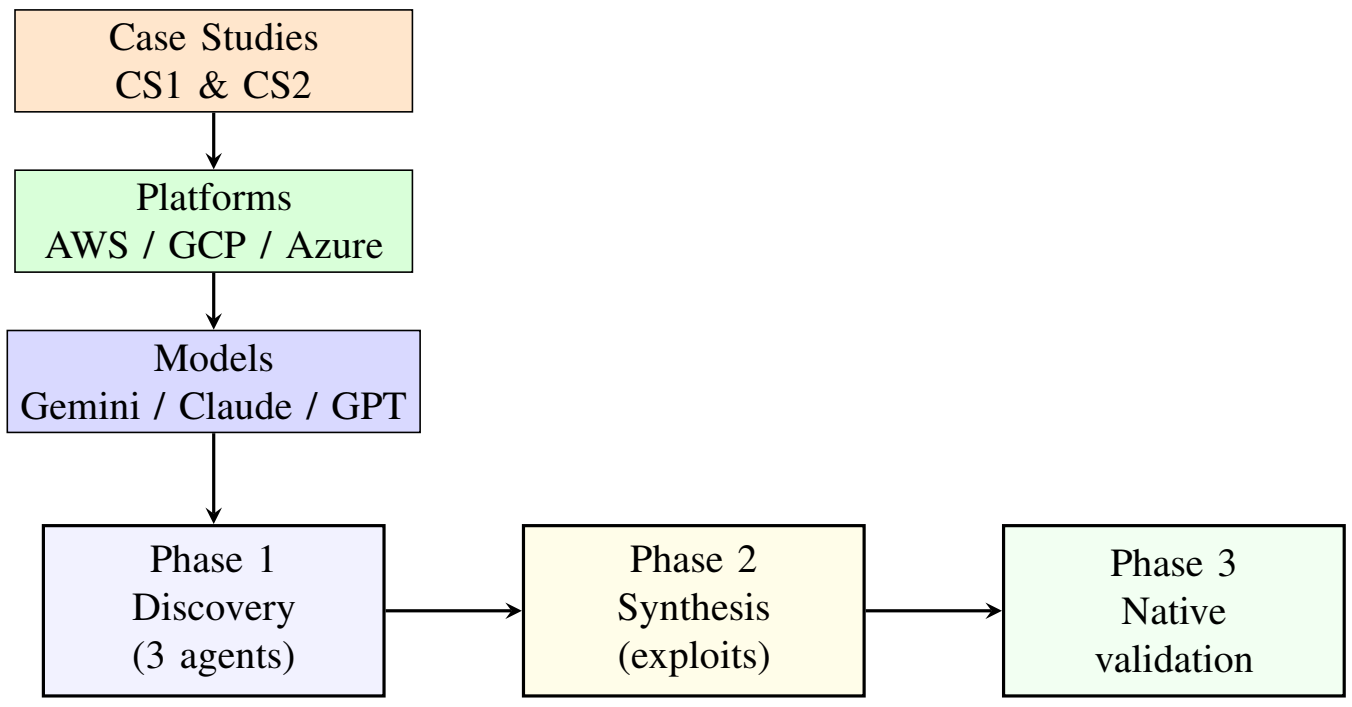

Our framework has three phases: multi-agent discovery, exploit synthesis, and empirical validation (Fig.~\ref{fig:methodology}).

\subsection{Three-Phase Validation}

Phase 1 (Multi-Agent Discovery): Three specialist agents (Application Security (AppSec), Infrastructure, Supply Chain) analyze baseline threat data to discover additional threats through cross-resource reasoning~\cite{chen2025agenticcyber}. Each agent receives platform metadata (IaC templates, network topology, identity and access management (IAM) policies) and outputs threats in JavaScript Object Notation (JSON) with severity, affected resources, and attack rationale. Agent specializations:

\begin{itemize}
\item \textbf{AppSec Agent:} Focuses on application-layer vulnerabilities, authentication/authorization bypass, injection attacks, and the security of web services reached over the Hypertext Transfer Protocol (HTTP). Analyzes exposed endpoints, application programming interface (API) configurations, and application-to-application trust boundaries. Specialized in Open Worldwide Application Security Project (OWASP) Top 10 threats adapted for cloud-native architectures.
\item \textbf{Infrastructure Agent:} Examines platform-level misconfigurations including IAM permission boundaries, network segmentation, encryption settings, and cloud service configurations. Analyzes cross-resource attack paths exploiting platform-specific features (AWS IAM role chaining, GCP service accounts, Azure managed identities).
\item \textbf{Supply Chain Agent:} Investigates CI/CD pipeline security, container image integrity, artifact repository access, and build-to-production trust boundaries~\cite{pan2024cicd}. Analyzes deployment automation, secret management in pipelines, and supply chain attack vectors including dependency confusion and malicious image injection.
\end{itemize}

The discovering model also assigns a severity rating to each threat. All three models received identical agent configurations (full prompts in the released artifact); no model-specific prompting or tuning was applied. Before validation, threats reported by more than one agent, run, model, or platform description within the same case study were deduplicated so that each distinct threat was tested only once. Descriptions are lowercased and whitespace-normalized, then compared pairwise and merged greedily above a similarity threshold. Both case studies use a single metric and threshold, the Jaccard overlap of the two word sets~\cite{pydifflib} at 0.24, selected as described in Section~\ref{sec:threat-set} by a coherence criterion the metric cannot see rather than tuned per case study. Section~\ref{sec:threat-set} reports a sweep of this threshold and its effect on the resulting threat set. Threats reported by more than one model form the consensus set analyzed in Section~\ref{sec:key-findings}.

\vspace{-0.1cm}
\begin{algorithm}[H]
\caption{Exploit Synthesis and Empirical Validation}
\label{alg:pipeline}
\small
\begin{algorithmic}[1]
\STATE \textbf{Part A --- Investigation.} \textbf{In:} threat $T$. \textbf{Out:} proof-of-concept commands $E$ and their transcripts
\STATE Parse attack vector $V$, components $C$, chain $S=\{s_1,\dots,s_n\}$ from $T$
\FOR{each step $s_i \in S$}
    \STATE Claude Sonnet~5 investigates the live deployment and runs non-destructive PoCs: boto3/gcloud/az for API, bash/curl for shell/HTTP; record transcripts
\ENDFOR
\STATE \textbf{Part B --- Validation.} \textbf{In:} $E$, transcripts, $T$. \textbf{Out:} class $\in$\{TP, FP, CONDITIONAL, INDETERMINATE\}, evidence
\STATE Deploy $T$'s case study natively on the cloud it was discovered against (Terraform/az CLI), as defined apart from operational fixes; pass a two-pass health gate
\STATE Run the PoCs from the attacker context below; capture control-plane responses, configs, probes, and PoC transcripts
\IF{the mechanism $T$ names is confirmed by a live PoC/artifact} \STATE class = TP
\ELSIF{a central claim is contradicted and no equivalent outcome is shown} \STATE class = FP
\ELSIF{the named mechanism is refuted but an equivalent outcome is shown via another real path} \STATE class = CONDITIONAL
\ELSE \STATE class = INDETERMINATE \textit{(discovery input wrong, or mechanism depends on a flow the testbed lacks)}
\ENDIF
\STATE Claude Opus~4.8 re-checks each verdict against the recorded evidence; an expert adjudicates; archive all artifacts
\end{algorithmic}
\end{algorithm}

Phase 2 (Exploit Synthesis): Algorithm~\ref{alg:pipeline} (Part A) details the procedure. For each threat, Claude Sonnet~5~\cite{anthropic2026sonnet5} parses the attack steps from the threat description, then investigates the live deployment and writes and runs proof-of-concept (PoC) commands, using the cloud providers' tools (boto3, the gcloud command-line interface, or the Azure command-line interface) for control-plane steps and curl or Requests for HTTP steps, recording a transcript of each. The PoCs are non-destructive: they probe and demonstrate access but do not damage or disrupt the target, although a few wrote benign test data to it. Unlike automatic exploit generation for binaries~\cite{avgerinos2011aeg,cha2012mayhem}, which searches program paths for memory-corruption exploits, our synthesis targets configuration and access-control weaknesses reachable through cloud APIs and service endpoints.

Phase 3 (Empirical Validation): Algorithm~\ref{alg:pipeline} (Part B) details the validation procedure. Each case study is deployed via Terraform (AWS and GCP) or the Azure command-line interface (Azure) on the same provider the models analyzed during discovery, as defined, with only the operational fixes needed to pass the health gate (instance resizing, image substitutions for unavailable or non-starting components); no change targeted a specific threat, and the configuration as validated is captured in the released evidence. Azure was additionally consolidated as noted in Section~\ref{sec:case-studies}. Every threat is exercised \emph{natively} on the provider it was discovered against, across six live deployments (two case studies on each of AWS, GCP, and Azure); before any evidence is captured, each deployment must pass a two-pass health gate confirming every service is operational.

\textit{Attacker context.} Evidence was gathered from three vantage points: unauthenticated requests to what each deployment exposes to the internet; host-level access through the operator's SSH session, standing in for an attacker who already holds a foothold (noted in each affected verdict); and read-only control-plane queries with the account owner's credentials to confirm configuration such as IAM policies, firewall rules and bucket policies.

Executing the exploits captures four kinds of evidence: cloud control-plane responses (IAM policies, instance metadata, and firewall/NSG and role queries on each provider), configuration files pulled from the hosts, service-endpoint probes, and literal proof-of-concept transcripts. From this evidence each threat is a true positive (TP) when the mechanism it names is confirmed, a false positive (FP) when a central claim is contradicted and no equivalent outcome is shown, conditional (CONDITIONAL) when the named mechanism is refuted but an equivalent impact is reached by a different real path (or some of its required facts are confirmed and others refuted), or indeterminate (INDETERMINATE) when the deciding part cannot be fairly tested because of the experiment rather than the model (detailed below). Two rules keep this conservative: a threat is FP only when a central claim is contradicted by live evidence and no equivalent outcome is shown, and every verdict must cite a captured artifact that contains the deciding fact.

\textit{Model usage, verification, and adjudication.} Live exploitation was performed by Claude Sonnet~5~\cite{anthropic2026sonnet5}, which executed the PoCs and recorded each verdict. The author authorized and supervised each validation run; exploits were limited to non-destructive actions against author-owned disposable testbeds. Two newer models, Claude Sonnet~5.5 and Claude Opus~5.5~\cite{anthropic2026opus55}, were repeatedly stopped by safety classifiers on the intentionally vulnerable testbeds and could not complete the live step---itself a methodological finding, since live validation of such testbeds currently depends on specific model versions. We therefore used Claude Opus~4.8~\cite{anthropic2026opus48} in a role it could perform, independently re-reviewing all 233 verdicts against the recorded evidence: it found no verdict wrong on the merits, though citation gaps on a few verdicts were queued for the expert.

\textit{Indeterminate rule.} By author decision, a refuted threat was relabeled indeterminate only when the refutation was caused by the experiment rather than the model: the discovery input contained contaminated data, or the threat's mechanism is a flow the case study is described as having but the testbed does not implement (Section~\ref{sec:case-studies}). This relabeled 37 threats; counting them as false positives would give 58.7\% precision. A security expert, a security team lead with over ten years of experience, then adjudicated a queue of all 61 conditional verdicts, a stratified sample of true and false positives from each deployment, and verdicts flagged in review (138 in total). The expert agreed with 97 (Cohen's $\kappa=0.60$); disagreements were reconciled against the written rules, and the final decisions, which changed 22 verdicts (chiefly 12 conditionals whose deciding link is a flow the testbed lacks, now indeterminate), give the labels reported in Section~\ref{sec:results}. Machine verdicts and expert decisions are stored separately, no machine verdict is overwritten, and both---with the full per-threat evidence bundle---are released with the artifact.

\section{Evaluation}

\subsection{Experimental Design}

We compare three commercial foundation models, one from each of three major model families, across three cloud platforms using a consistent three-agent architecture across two case studies. Models evaluated: Gemini 3.8 Flash~\cite{gemini2026flash} (Google), Claude Haiku 4.5~\cite{anthropic2025haiku} (Anthropic), and GPT-5.6 Luna~\cite{openai2026gpt56} (OpenAI). Each model was run five times per platform and case study; the reported threats are the unique threats accumulated across the five runs after deduplication.

\subsection{Case Studies}
\label{sec:case-studies}

Both case studies are taken from our prior work on semi-automated threat modeling~\cite{pecka2026semiautomated}, which describes them in full; we summarize them here. Both were built as testbeds: each of the 17 threat types was deliberately introduced into each deployment using the platform's native security constructs (for example, unauthenticated external access on AWS is a security group allowing ingress from 0.0.0.0/0 to the Jenkins port), so that the baseline analyzers have known threats to detect. In both case studies every service runs as a Docker container on virtual machines (VMs), and both span three trust zones whose boundaries are crossed by code, container images, and models on their way to production.

Case Study 1 (CS1) is an eight-component ML supply chain across three trust zones: code and training scripts in Gitea; Jenkins building images and running training jobs; a Docker Registry and MLflow holding images and model artifacts; and, in production, a Node.js application over PostgreSQL calling a Flask fraud-detection inference service. Case Study 2 (CS2) is a 20-component container registry built on Project Quay: images are pushed through an NGINX proxy to two Quay instances, layers stored in MinIO and metadata in PostgreSQL, with Redis caching, Clair vulnerability scanning, an LLM analysis service, and Jenkins, over supporting cloud resources (VPC, security groups, S3 buckets, IAM roles).

On AWS, CS1 runs on three small EC2 instances (t3.micro/t3.small, one per zone) and CS2 on seven, giving CS2 about $2.5\times$ the components of CS1; the same architectures were deployed on GCP (Compute Engine, Cloud Storage, service accounts) and Azure (VMs, Blob Storage), with Azure consolidated onto two VMs per case study under account quota, with some services replaced by stand-ins. AWS and GCP deployments are defined in Terraform and Azure through the Azure CLI, and configuration was extracted through the platform API as in~\cite{pecka2026semiautomated}. The two case studies thus differ in both architecture and scale.

The testbeds deploy every component and the planted misconfigurations but not the data flows between them: no source repositories or CI jobs are provisioned, the inference service returns a fixed rule rather than loading a model, and no workload consumes the registries. Threats whose mechanism depends on these flows cannot be exercised and are labeled indeterminate.

\subsection{Baseline Threat Detection}

Baseline threats were initially identified using the semi-automated methodology of Pecka et al.~\cite{pecka2026semiautomated}: (1) deploy architecture, (2) extract configuration via cloud APIs, (3) capture network flows, (4) apply deterministic analyzers detecting 17 threat types based on industry security benchmarks~\cite{cis2024benchmarks} --- ten general (unauthenticated access, excessive IAM permissions, unencrypted cross-boundary flows, missing network segmentation, anonymous registry access, CI/CD-to-production access, missing audit logging, exposed secrets, overly broad source control, and missing resource limits) and seven ML/AI-specific (unauthorized model access, training-data poisoning, model theft via API, unencrypted model transfer, ML-pipeline-to-production access, missing model provenance, and exposed model-metadata endpoints) --- (5) validate 100\% type coverage. These baseline threats (CS1 166: AWS 49, GCP 53, Azure 64; CS2 192: AWS 75, GCP 53, Azure 64) served as input context for AI models but are not included in the 233 validated threats reported in this paper. That prior work also compared its deterministic analyzers against six external scanners on these same testbeds: the best, Prowler and Kubescape, detected 8 of the 17 threat types, Checkov 3, and Trivy and KubeLinter 1 each, and none detected any of the seven ML/AI-specific types~\cite{pecka2026semiautomated}. The 233 AI-discovered threats lie outside that baseline by construction: cross-component attack paths found by multi-agent reasoning that no evaluated scanner or deterministic analyzer reports.

\subsection{Threat Set Construction and Labeling}
\label{sec:threat-set}

The validated set was assembled in four steps. First, across five runs of each model on each platform, the agents reported 288 raw threats (CS1 158, CS2 130). Second, we removed 16 CS1 records that reproduced the output schema rather than stating a threat (for example the literal strings ``Clear description of the cross-resource threat'' and ``Detailed threat description'', recorded with the severity field left as the enumeration \texttt{Critical|High|Medium|Low}). Third, deduplication merged restatements of the same threat, leaving 235 distinct threats (CS1 131, CS2 104). Finally we removed two CS1 records describing components of the other case study, which had been tested against the wrong deployment. The remaining 233 threats (CS1 129, CS2 104) are the ones analyzed in Section~\ref{sec:results}. Eleven threats reported on more than one provider were merged by deduplication and validated on the first provider that reported them.

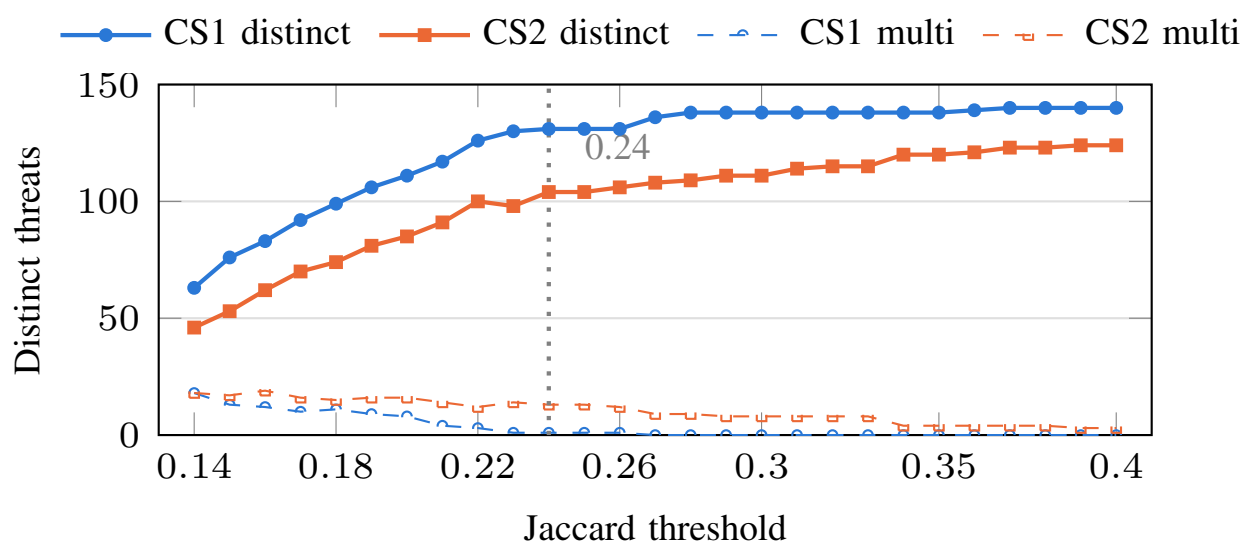
\begin{figure}[t]
\centering
\begin{tikzpicture}
\begin{axis}[
    width=0.92\columnwidth, height=3.9cm,
    xmin=0.13, xmax=0.41, ymin=0, ymax=150,
    xlabel={Jaccard threshold}, ylabel={Distinct threats},
    xlabel style={font=\scriptsize}, ylabel style={font=\scriptsize},
    xticklabel style={font=\scriptsize}, yticklabel style={font=\scriptsize},
    xtick={0.14,0.18,0.22,0.26,0.30,0.35,0.40},
    ymajorgrids, grid style={gray!25},
    legend style={at={(0.5,1.03)}, anchor=south, legend columns=4,
                  font=\scriptsize, draw=none, /tikz/every even column/.append style={column sep=3pt}},
]
\addplot[tpblue, mark=*, mark size=0.9pt, thick] coordinates
  {(0.140,63)(0.150,76)(0.160,83)(0.170,92)(0.180,99)(0.190,106)(0.200,111)(0.210,117)(0.220,126)(0.230,130)(0.240,131)(0.250,131)(0.260,131)(0.270,136)(0.280,138)(0.290,138)(0.300,138)(0.310,138)(0.320,138)(0.330,138)(0.340,138)(0.350,138)(0.360,139)(0.370,140)(0.380,140)(0.390,140)(0.400,140)};
\addplot[fporange, mark=square*, mark size=0.9pt, thick] coordinates
  {(0.140,46)(0.150,53)(0.160,62)(0.170,70)(0.180,74)(0.190,81)(0.200,85)(0.210,91)(0.220,100)(0.230,98)(0.240,104)(0.250,104)(0.260,106)(0.270,108)(0.280,109)(0.290,111)(0.300,111)(0.310,114)(0.320,115)(0.330,115)(0.340,120)(0.350,120)(0.360,121)(0.370,123)(0.380,123)(0.390,124)(0.400,124)};
\addplot[tpblue, mark=o, mark size=0.9pt, dashed] coordinates
  {(0.140,18)(0.150,13)(0.160,12)(0.170,10)(0.180,11)(0.190,9)(0.200,8)(0.210,4)(0.220,3)(0.230,1)(0.240,1)(0.250,1)(0.260,1)(0.270,0)(0.280,0)(0.290,0)(0.300,0)(0.310,0)(0.320,0)(0.330,0)(0.340,0)(0.350,0)(0.360,0)(0.370,0)(0.380,0)(0.390,0)(0.400,0)};
\addplot[fporange, mark=square, mark size=0.9pt, dashed] coordinates
  {(0.140,18)(0.150,17)(0.160,19)(0.170,16)(0.180,15)(0.190,16)(0.200,16)(0.210,14)(0.220,12)(0.230,14)(0.240,13)(0.250,13)(0.260,12)(0.270,9)(0.280,9)(0.290,8)(0.300,8)(0.310,8)(0.320,8)(0.330,8)(0.340,4)(0.350,4)(0.360,4)(0.370,4)(0.380,4)(0.390,3)(0.400,3)};
\legend{CS1 distinct, CS2 distinct, CS1 multi, CS2 multi}
\draw[gray,dotted,thick] (axis cs:0.24,0) -- (axis cs:0.24,150);
\node[anchor=south west,font=\scriptsize,gray] at (axis cs:0.245,110) {0.24};
\end{axis}
\end{tikzpicture}
\caption{Deduplication threshold sweep (selected Jaccard metric). Solid: distinct threats after merging; dashed: threats reported by more than one model. The dotted line marks the 0.24 operating point; the multi-model count is highly threshold-sensitive (18$\to$0 in CS1, 1 at the operating point).}
\label{fig:dedup-sensitivity}
\end{figure}

Deduplication was treated as a measurement choice rather than a tuned constant. We compared three similarity metrics (the difflib sequence ratio, the Jaccard overlap of the word sets, and their mean), scoring each at matched group counts by a criterion the metrics cannot see: the mean pairwise overlap of the \texttt{affected\_resources} field within each merged group, since genuine restatements of one threat should name the same resources. Jaccard dominated, beating the sequence ratio on 25 of 29 matched operating points by a mean of 0.064 coherence (two-sided sign test, $p=0.0001$); the sequence ratio, used by earlier versions of this pipeline, merged on surface vocabulary and in one case fused three distinct threats into one. We therefore apply Jaccard at 0.24 to both case studies, the joint coherence optimum (Fig.~\ref{fig:dedup-sensitivity}); neighboring thresholds lie within the bootstrap intervals, so the metric choice matters more than the exact value. One consequence carries into Finding~1: the number of threats recognized as multi-model falls from 18 to 0 in CS1 across the sweep (1 at the operating point), so a looser metric manufactures apparent agreement, and any description-similarity consensus signal inherits that sensitivity.

These 233 threats are the models' claims, each of which may be right or wrong; they are not a ground-truth inventory. Ground truth exists at two levels. For the baseline, it is known by construction: the 17 threat types were deliberately introduced into each deployment, so the deterministic analyzers' output can be scored against them~\cite{pecka2026semiautomated}. The AI-discovered threats lie, by design, outside this baseline, so no reference list exists for them. Instead, the label of each threat is established individually from the live system by the Phase 3 procedure: a TP or FP is recorded only when evidence captured from the running deployment confirms or contradicts the mechanism the threat names, a CONDITIONAL label records that the named mechanism was refuted but an equivalent impact was demonstrated through a different real path, every verdict is independently re-checked against its evidence and contested ones are adjudicated by a security expert, and threats that cannot be fairly tested (contaminated discovery input, or a mechanism that depends on a flow the testbed lacks) are labeled INDETERMINATE rather than guessed. These labels are therefore ground truth for precision (whether a reported threat is real) but not for recall, because threats that no model reported cannot be counted.

\section{Results and Analysis}
\label{sec:results}

\subsection{Threat Discovery}

The 233 threats are distributed unevenly across models and platforms (each counted once, under the first reporting model in the order below). Gemini 3.8 Flash and Haiku 4.5 contributed similar numbers (100 and 88) and GPT-5.6 Luna about half as many (45); by platform description, GCP yielded the most (85), AWS 80, and Azure the fewest (68). Fourteen threats (6.0\%) were reported by more than one model, one in CS1 and thirteen in CS2; they form the consensus set analyzed in Finding~1. The full model$\times$platform$\times$case-study breakdown is in the released artifact.

\subsection{Empirical Validation}

\begin{table}[t]
\centering
\caption{Cumulative Validation Results Across All Threats}
\label{tab:case-comparison}
\small\setlength{\tabcolsep}{4pt}
\begin{tabular}{|l|c|c|c|}
\hline
\textbf{Metric} & \textbf{CS1} & \textbf{CS2} & \textbf{Overall} \\
\hline
Components & 8 & 20 & -- \\
Trust Zones & 3 & 3 & -- \\
Validated Threats & 129 & 104 & 233 \\
\hline
\textbf{TP Rate} & \textbf{42.6\%} & \textbf{46.2\%} & \textbf{44.2\%} \\
 & (55/129) & (48/104) & (103/233) \\
\textbf{FP Rate} & \textbf{12.4\%} & \textbf{17.3\%} & \textbf{14.6\%} \\
 & (16/129) & (18/104) & (34/233) \\
Conditional & 20.9\% (27) & 16.3\% (17) & 18.9\% (44) \\
Indeterminate & 24.0\% (31) & 20.2\% (21) & 22.3\% (52) \\
\textbf{Precision} & \textbf{77.5\%} & \textbf{72.7\%} & \textbf{75.2\%} \\
 & (55/71) & (48/66) & (103/137) \\
\textit{95\% CI} & [66.5, 85.6] & [61.0, 82.0] & [67.3, 81.7] \\
\hline
\multicolumn{4}{|c|}{\textbf{By Model Consensus}} \\
\hline
Multi-Model TP & \multicolumn{3}{c|}{42.9\% (6/14)} \\
Single-Model TP & \multicolumn{3}{c|}{44.3\% (97/219)} \\
Multi-Model Precision & \multicolumn{3}{c|}{60.0\% (6/10)} \\
Single-Model Precision & \multicolumn{3}{c|}{76.4\% (97/127)} \\
\hline
\multicolumn{4}{|c|}{\textbf{By Severity (precision)}} \\
\hline
Critical & \multicolumn{3}{c|}{72.9\% (35/48)} \\
High & \multicolumn{3}{c|}{73.4\% (47/64)} \\
Medium & \multicolumn{3}{c|}{84.0\% (21/25)} \\
\hline
\end{tabular}

\vspace{2pt}
\begin{minipage}{0.95\columnwidth}
\footnotesize
Note: TP, FP, conditional, and indeterminate rates are over all threats in each column. Precision is TP/(TP+FP), over resolved (TP+FP) threats only; conditional and indeterminate threats are excluded from the precision denominator. Intervals are 95\% Wilson score intervals.
\end{minipage}
\end{table}

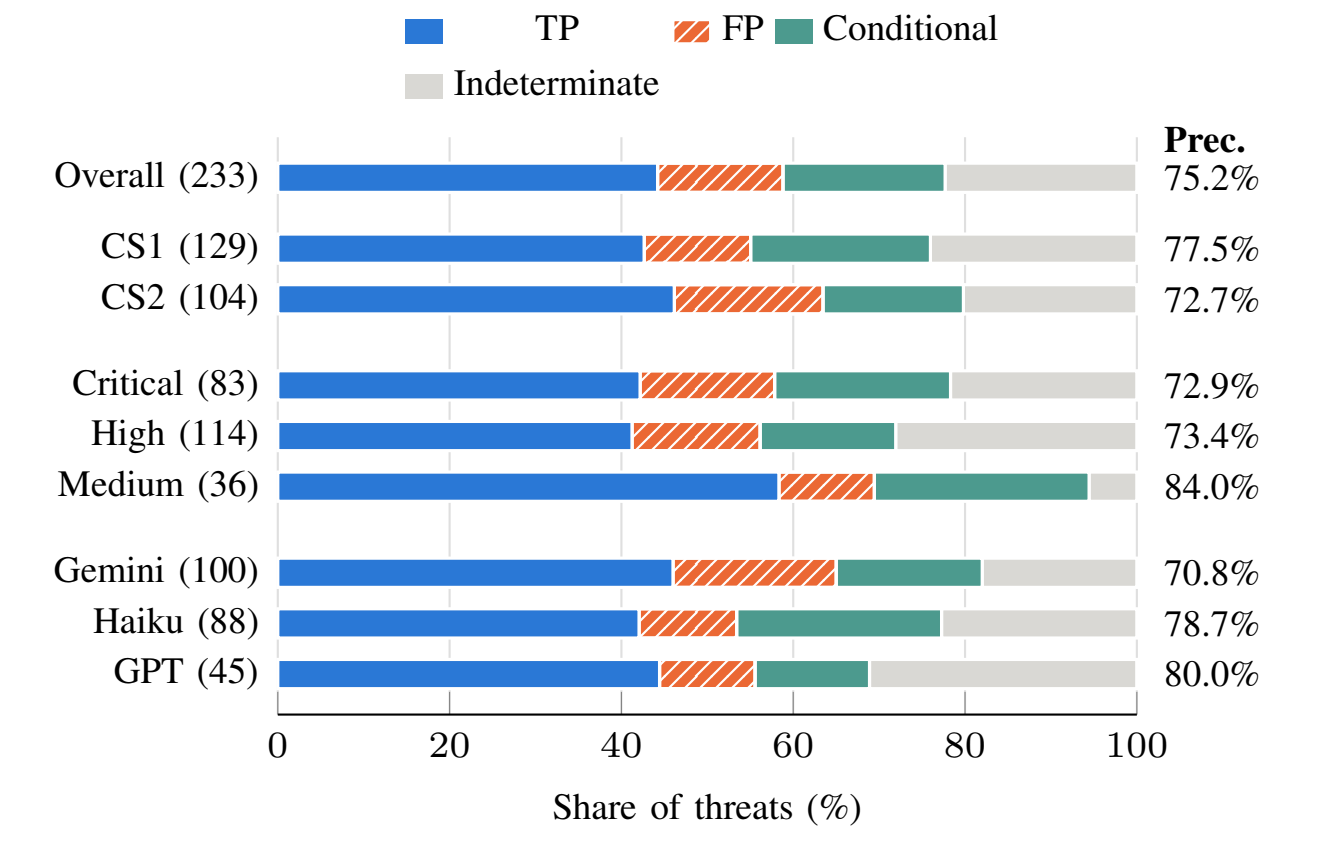
\begin{figure}[t]
\centering
\begin{tikzpicture}
\begin{axis}[
    width=0.82\columnwidth, height=5.4cm,
    xbar stacked, bar width=5.5pt,
    xmin=0, xmax=100, ymin=0.4, ymax=11.8,
    clip=false, axis lines*=left,
    ytick={11,9.6,8.6,6.9,5.9,4.9,3.2,2.2,1.2},
    yticklabels={Overall (233),CS1 (129),CS2 (104),Critical (83),High (114),Medium (36),Gemini (100),Haiku (88),GPT (45)},
    yticklabel style={font=\scriptsize},
    xticklabel style={font=\scriptsize},
    xlabel={Share of threats (\%)}, xlabel style={font=\scriptsize},
    xmajorgrids, grid style={gray!25},
    y axis line style={draw=none}, ytick style={draw=none},
    legend style={at={(0.5,1.02)}, anchor=south, legend columns=3, font=\scriptsize, draw=none},
    legend image code/.code={\draw[#1,draw=none] (0cm,-0.08cm) rectangle (0.25cm,0.08cm);},
]
\addplot[fill=tpblue, draw=white, line width=0.6pt] coordinates {(44.21,11) (42.64,9.6) (46.15,8.6) (42.17,6.9) (41.23,5.9) (58.33,4.9) (46.00,3.2) (42.05,2.2) (44.44,1.2)};
\addplot[fill=fporange, draw=white, line width=0.6pt, postaction={pattern=north east lines, pattern color=white}] coordinates {(14.59,11) (12.40,9.6) (17.31,8.6) (15.66,6.9) (14.91,5.9) (11.11,4.9) (19.00,3.2) (11.36,2.2) (11.11,1.2)};
\addplot[fill=condteal, draw=white, line width=0.6pt] coordinates {(18.88,11) (20.93,9.6) (16.35,8.6) (20.48,6.9) (15.79,5.9) (25.00,4.9) (17.00,3.2) (23.86,2.2) (13.33,1.2)};
\addplot[fill=indgray, draw=white, line width=0.6pt] coordinates {(22.32,11) (24.03,9.6) (20.19,8.6) (21.69,6.9) (28.07,5.9) (5.56,4.9) (18.00,3.2) (22.73,2.2) (31.11,1.2)};
\legend{TP, FP, Conditional, Indeterminate}
\node[anchor=west,font=\scriptsize] at (axis cs:101,11) {75.2\%};
\node[anchor=west,font=\scriptsize] at (axis cs:101,9.6) {77.5\%};
\node[anchor=west,font=\scriptsize] at (axis cs:101,8.6) {72.7\%};
\node[anchor=west,font=\scriptsize] at (axis cs:101,6.9) {72.9\%};
\node[anchor=west,font=\scriptsize] at (axis cs:101,5.9) {73.4\%};
\node[anchor=west,font=\scriptsize] at (axis cs:101,4.9) {84.0\%};
\node[anchor=west,font=\scriptsize] at (axis cs:101,3.2) {70.8\%};
\node[anchor=west,font=\scriptsize] at (axis cs:101,2.2) {78.7\%};
\node[anchor=west,font=\scriptsize] at (axis cs:101,1.2) {80.0\%};
\node[anchor=west,font=\scriptsize\bfseries] at (axis cs:101,11.75) {Prec.};
\end{axis}
\end{tikzpicture}
\caption{Validation outcomes by case study, model-assigned severity, and discovering model (n in parentheses). Bars show the share of threats confirmed (TP), refuted (FP), conditional (the named mechanism refuted but an equivalent impact shown through another real path), or indeterminate (cannot be fairly tested: contaminated input or a flow the testbed lacks); the right column gives precision, TP/(TP+FP), over resolved threats. Model rows count each threat under its first reporting model.}
\label{fig:outcomes}
\end{figure}

We tested all 233 threats natively, each against the provider it was discovered on, using non-destructive live exploitation across six deployments. The captured evidence resolved 137 of them (58.8\%) as true or false positives: 103 were confirmed (TP) and 34 refuted (FP), a precision of 75.2\% (95\% CI [67.3, 81.7]). A further 44 (18.9\%) were conditional: the specific mechanism the threat named was refuted, but an equivalent impact was demonstrated through a different real path, so the threat is neither a clean confirmation nor a clean refutation. The remaining 52 (22.3\%) were indeterminate because they could not be fairly tested: the discovery input was wrong, or the threat's mechanism depends on a flow the testbed does not implement (for example a CI job that was never provisioned), which can be settled only by redeploying a fuller architecture and rediscovering, not by further testing of this one. Because conditional and indeterminate threats count as neither TP nor FP, the whole-set rates (44.2\% TP and 14.6\% FP) are not directly comparable to precision. Table~\ref{tab:case-comparison} and Fig.~\ref{fig:outcomes} break these outcomes down by case study, severity, consensus, and model.

The two case studies produced similar outcome mixes (precision 77.5\% for CS1, 72.7\% for CS2), and confirmed threats span the same high-impact classes in both: Docker-socket container escape, wildcard IAM roles whose credentials are retrievable from unauthenticated instance metadata, plaintext credentials in startup configuration, unauthenticated registries and inference endpoints, publicly readable object storage, internet-reachable databases, SSH open to the internet, and absent network segmentation and logging. False positives were not confined to one case study (16 of 129 CS1 threats and 18 of 104 CS2 threats were refuted), as Finding~2 examines.

The discovering models assigned a severity to every one of the 233 threats: 83 Critical, 114 High, and 36 Medium. No model ever assigned Low, although it was offered in the output schema given to the agents, so the severity range evaluated here is Critical through Medium. Among resolved threats, precision did not rise with assigned severity (Critical 72.9\%, High 73.4\%, Medium 84.0\%, if anything the reverse), a pattern examined in Finding~3.

\subsection{Key Findings}
\label{sec:key-findings}

\subsubsection*{Finding 1: Multi-model discovery adds coverage, not agreement}
The three models found largely different threats: only 14 of 233 were reported by more than one model, and only 6 of the 103 confirmed threats. Counting each threat under every model that reported it, Gemini 3.8 Flash accounts for 46 confirmed threats, Haiku 4.5 for 43, and GPT-5.6 Luna for 20; together they yield 103, 2.2$\times$ the best single model, at 75.2\% precision against 70.8--80.0\% for each model alone. Two cautions apply: without a ground-truth inventory this is a gain in confirmed findings, not a measured recall, and the combined set reflects three times as many runs as any one model (five per model and cell), so part of the gain may come from repeated sampling rather than model diversity alone.

Agreement between models, by contrast, was not a confidence signal. Of the 14 multi-model threats, 6 were confirmed (42.9\%), 4 were refuted, 1 was conditional, and 3 were indeterminate. Single-model threats were confirmed at 44.3\% (97/219). Agreement therefore did not raise the confirmation rate (two-sided Fisher exact test: $p=1.0$; cluster-robust by case-study$\times$provider cell, $p=0.84$), and among resolved threats it did not raise precision either: 60.0\% (6/10) for multi-model against 76.4\% (97/127) for single-model ($p=0.26$). Both point estimates favor single-model discovery, though with 14 consensus threats neither difference is distinguishable from noise. What we can say is that we find no evidence that agreement between models predicts whether a threat is real, and that four of the 34 false positives were themselves multi-model threats.

This result is also instrument-dependent: the consensus set ranges from 18 threats to 0 in CS1 across the threshold sweep, because looser matching merges distinct threats that different models phrased similarly into less coherent groups. Future work should define agreement semantically, for example by whether two threats resolve against the same captured evidence.

\subsubsection*{Finding 2: False positives assert access or state the system does not have}
False positives occurred in all six deployments (CS1: 3 on AWS, 9 on GCP, 4 on Azure; CS2: 5, 7, and 6). The CS1/CS2 difference is not significant (two-sided Fisher exact test, $p=0.56$; cluster-robust by case-study$\times$provider cell, $p=0.52$), nor is the difference across providers (AWS 8/80, GCP 16/85, Azure 10/68; $\chi^2$ $p=0.28$). What the 34 false positives share is not a location but a failure mode: they assert access or system state the deployment did not have. Of the 34, 18 over-claim access (11 permission, grant, or write paths the live system denies, and 7 that assume authentication is absent when it is enforced) and 16 invent system state (9 configuration, 4 capability, and 3 component or flow that no case-study design contains). Examples recur: on GCP, several threats treat the \texttt{cloud-platform} OAuth scope attached to an instance as if it conferred broad permissions, but the service account shared by dev and prod holds no project IAM role and a live API call returns HTTP 403; on AWS CS1, three claims of writable public buckets fail because the bucket policy is read-only and an anonymous PUT is refused with 403. These failure modes were assigned by the AI assistant after the verdicts were known, so the assignment is not blind; the per-threat table is released with the artifact. The modes are addressable: prompting models to cite the specific configuration artifact (the IAM policy document, the role's attached policies, the identity binding) that would support each privileged step or claimed component would ground each claim in observable state.

\subsubsection*{Finding 3: Models identify the weakness; what they get wrong is the mechanism}
Model-assigned severity was a weak indicator of correctness, and if anything ran the wrong way: resolved Critical threats had the lowest precision (72.9\%, 35/48), High 73.4\% (47/64), and Medium the highest (84.0\%, 21/25). The Critical--High difference is not significant (two-sided Fisher exact test, $p=1.00$), consistent with empirical research showing that severity scores poorly predict actual exploitation~\cite{wang2024secscore}. The stronger signal is in the 44 conditional threats (18.9\% of the set). In these the model correctly identified a real architectural weakness but misstated the mechanism by which it is exploited: it named default object-store credentials where the deployed store enforces no authentication at all, or a Jenkins-to-production pipeline that is not wired even though the same host is reachable by another path. The outcome the threat warned about is often real; the specific route it describes is not.

To test where this mechanism error falls without circularity, we coded every threat from its description alone, before consulting any verdict, as \emph{observable} (a single static property checkable by one configuration read or probe, such as an open port, a wildcard IAM role, or a public bucket; 80 threats) or \emph{behavioral} (requiring a prior compromise step, downstream propagation, or runtime behavior; 153). Observable threats resolved to TP or FP more often (72.5\% vs.\ 51.6\%; Fisher $p=0.002$, Mantel--Haenszel stratified by deployment $p=0.004$), though not in every deployment (reversed in CS2 Azure and GCP), and the difference persists when indeterminate threats are excluded (86.6\% vs.\ 69.3\%). They were also conditional less often (11.2\% vs.\ 22.9\%, Fisher $p=0.03$), though less consistently within deployments (stratified $p=0.06$), and their precision was higher (82.8\% vs.\ 69.6\%; Fisher $p=0.11$, stratified $p=0.04$), though not under the cluster-robust estimate ($p=0.19$). Reversing every borderline coding decision leaves these differences essentially unchanged. Taken together with the over-claimed access and invented state among the false positives (Finding~2), model error concentrates in the \emph{exploitation mechanism} of multi-step chains rather than in whether a weakness exists. The practical consequence is that an AI-discovered threat is more reliable as a lead to a genuine weakness than as a validated exploitation path, and teams should confirm the named mechanism against live configuration before acting on it as described.

\subsection{Practical Implications}
Three guidelines follow for teams using AI threat discovery. First, run several models for coverage: individual models were close in confirmation rate (42--46\%) and precision (71--80\%), so using more than one matters more than which one (Finding~1). Second, do not treat agreement between models as pre-validation. Third, triage by observability: threats that turn on directly observable configuration (IAM scope, security groups, public buckets, encryption) can be auto-verified with deterministic checks, whereas multi-step chains, which most often land conditional, should be validated empirically before teams act on them (Finding~3). Severity remains the right basis for remediation priority, not for judging whether a finding is real.

\section{Threats to Validity and Limitations}

\textit{Internal validity.} The case studies run on live AWS, GCP, and Azure infrastructure built from production open-source software (Quay, Clair, Jenkins, Gitea, MLflow, PostgreSQL, Redis, MinIO), but they are research testbeds with known threats deliberately introduced, so results may differ where production adds controls not captured in IaC, such as runtime security agents. Non-destructive live validation resolved 58.8\% of threats as TP or FP; the conditional (18.9\%) and indeterminate (22.3\%) labels handle the rest conservatively rather than forcing a binary verdict.

\textit{External validity.} Validation is native on all three providers --- each threat exercised on the cloud it was discovered against, across six deployments --- but covers only two microservice architectures; serverless, analytics pipelines, VM monoliths, hybrid multi-cloud, and smaller providers remain untested, and we draw no architecture-level conclusions (false-positive behavior did not differ significantly between the two we tested). The Azure deployments ran at reduced scale under account quota, and the testbeds omit the data flows between components (Section~\ref{sec:case-studies}), so threats depending on them are among the 52 indeterminate ones, which would need a fuller architecture deployed and rediscovered to settle rather than being model errors.

\textit{Construct validity.} A TP confirms that a threat's asserted conditions are present, which is weaker than demonstrating the full attack end to end. Conditional and indeterminate threats count as neither TP nor FP, so whole-set rates are not comparable to precision (computed over TP+FP only). Recall is not measured: with no ground-truth inventory, threats missed by all models cannot be counted. The observable/behavioral coding (Finding~3) was done by a single coder from descriptions alone. Severity is model-assigned, and application-level vulnerabilities (SQL injection, cross-site scripting, business-logic flaws) are underrepresented; future work should test AI-discovered application threats against vulnerable-by-design apps such as WebGoat and DVWA.

\textit{Selection and statistics.} Deduplication merged 37 of 272 non-placeholder records (13.6\%), preventing repeated testing of one threat but possibly collapsing legitimate variants; the metric was selected by the Section~\ref{sec:threat-set} procedure, which ranks the three metrics rather than proving an optimum. The consensus comparison (Finding~1) is not significant under a Fisher exact test or a cluster-robust estimate by case-study$\times$provider cell; because threats are nested within architectures, providers, and models, we report cluster-robust p-values alongside the Fisher values and treat the latter as optimistic.

\textit{Temporal validity and future work.} Results reflect Fall 2026 models (Gemini 3.8 Flash, Claude Haiku 4.5, GPT-5.6 Luna), so continuous re-evaluation is needed. Validating at discovery time via cloud-API monitoring would shorten the discovery-to-confirmation gap; redeploying a fuller architecture with the missing data flows would resolve the indeterminate cases, and destructive testing in disposable environments would settle the conditional ones; and a human-analyst comparison plus Low-severity threats would complete the picture.

\section{Conclusion}

This paper presented an empirical validation of AI-discovered cloud infrastructure threats, addressing open questions about accuracy, reliability, and systematic failure modes. Through native live validation of 233 threats on the cloud each was discovered against, across three providers and two architectures, we found 75.2\% precision among the threats that non-destructive testing resolved as true or false positives, supporting AI-assisted threat discovery as a practical input to security workflows when combined with empirical validation.

Three key insights emerge from our validation. First, multi-model discovery pays off through breadth: three models together more than doubled the confirmed threats of the best single model at comparable precision, whereas agreement between models raised neither the confirmation rate nor precision. Second, false positives were not tied to an architecture but asserted access or system state the deployment did not have, such as treating a cloud OAuth scope as an effective grant. Third, neither severity nor architecture predicted whether a resolved threat was correct; instead, multi-step behavioral threats, identified before validation, were the ones most often conditional, suggesting that models more often misidentify the exploitation mechanism than the underlying weakness. Our methodology and results establish a foundation for integrating AI security tools into cloud security operations with quantified confidence levels.

\textbf{Data availability.} The validation framework, complete results (233 threats), deduplication sweep and metric selection, per-threat evidence bundle, expert decisions, and threat taxonomy are released at \url{https://github.com/npecka/multi-agent-threat-verification}.

\bibliographystyle{IEEEtran}
\bibliography{main}

@article{pecka2026semiautomated,
  author={Pecka, Nicholas and Ben Othmane, Lotfi and Bhargava, Bharat and Bryce, Ren{\'e}e},
  title={Semi-Automated Threat Modeling of Cloud-Based Systems Through Extracting Software Architecture from Configuration and Network Flow},
  journal={IEEE Transactions on Services Computing},
  pages={1--13},
  year={2026},
  doi={10.1109/TSC.2026.3724971},
  note={Early Access}
}

@misc{Crossman2025Auspex,
  title={{Auspex}: Building Threat Modeling Tradecraft into an Artificial Intelligence-based Copilot},
  author={Crossman, Andrew and Plummer, Andrew R. and Sekharudu, Chandra and Warrier, Deepak and Yekrangian, Mohammad},
  year={2025},
  eprint={2503.09586},
  archivePrefix={arXiv},
  primaryClass={cs.CR},
  url={https://arxiv.org/abs/2503.09586}
}

@inproceedings{tramer2016stealing,
  title={Stealing Machine Learning Models via Prediction {APIs}},
  author={Tram{\`e}r, Florian and Zhang, Fan and Juels, Ari and Reiter, Michael K. and Ristenpart, Thomas},
  booktitle={25th USENIX Security Symposium (USENIX Security 16)},
  pages={601--618},
  year={2016}
}

@misc{gu2017badnets,
  title={{BadNets}: Identifying Vulnerabilities in the Machine Learning Model Supply Chain},
  author={Gu, Tianyu and Dolan-Gavitt, Brendan and Garg, Siddharth},
  year={2017},
  eprint={1708.06733},
  archivePrefix={arXiv},
  primaryClass={cs.CR},
  url={https://arxiv.org/abs/1708.06733}
}

@inproceedings{shokri2017membership,
  title={Membership Inference Attacks Against Machine Learning Models},
  author={Shokri, Reza and Stronati, Marco and Song, Congzheng and Shmatikov, Vitaly},
  booktitle={2017 IEEE Symposium on Security and Privacy (SP)},
  pages={3--18},
  year={2017},
  doi={10.1109/SP.2017.41}
}

@misc{deng2023pentestgpt,
  title={{PentestGPT}: An {LLM}-empowered Automatic Penetration Testing Tool},
  author={Deng, Gelei and Liu, Yi and Mayoral-Vilches, V{\'\i}ctor and Liu, Peng and Li, Yuekang and Xu, Yuan and Zhang, Tianwei and Liu, Yang and Pinzger, Martin and Rass, Stefan},
  year={2023},
  eprint={2308.06782},
  archivePrefix={arXiv},
  primaryClass={cs.SE},
  url={https://arxiv.org/abs/2308.06782}
}

@misc{traust2024,
  author={{Red Hat Hybrid Platforms Team}},
  title={Traust: Workflow Engine for Automated Security Assessment},
  year={2026},
  howpublished={\url{https://github.com/traust-security/traust}},
  note={Apache License 2.0}
}

@misc{gemini2026flash,
  author={{Google DeepMind}},
  title={{Gemini 3.8 Flash Model Card}},
  year={2026},
  month=sep,
  howpublished={\url{https://deepmind.google/models/model-cards/gemini-3-8-flash/}},
  note={Accessed: 2026-10-03}
}

@misc{kumar2024stride,
  title={Asset-centric Threat Modeling for {AI}-based Systems},
  author={von der Assen, Jan and Sharif, Jamo and Feng, Chao and Killer, Christian and Bovet, G{\'e}r{\^o}me and Stiller, Burkhard},
  year={2024},
  eprint={2403.06512},
  archivePrefix={arXiv},
  primaryClass={cs.CR},
  url={https://arxiv.org/abs/2403.06512}
}

@inproceedings{avgerinos2011aeg,
  title={{AEG}: Automatic Exploit Generation},
  author={Avgerinos, Thanassis and Cha, Sang Kil and Hao, Brent Lim Tze and Brumley, David},
  booktitle={Proceedings of the Network and Distributed System Security Symposium (NDSS)},
  year={2011},
  month={February},
  address={San Diego, CA},
  url={https://www.ndss-symposium.org/ndss2011/aeg-automatic-exploit-generation}
}

@inproceedings{cha2012mayhem,
  title={Unleashing {MAYHEM} on Binary Code},
  author={Cha, Sang Kil and Avgerinos, Thanassis and Rebert, Alexandre and Brumley, David},
  booktitle={2012 IEEE Symposium on Security and Privacy},
  pages={380--394},
  year={2012},
  organization={IEEE},
  doi={10.1109/SP.2012.31}
}

@misc{zhang2023hallucination,
  title={Siren's Song in the {AI} Ocean: A Survey on Hallucination in Large Language Models},
  author={Zhang, Yue and Li, Yafu and Cui, Leyang and Cai, Deng and Liu, Lemao and Fu, Tingchen and Huang, Xinting and Zhao, Enbo and Zhang, Yu and Xu, Chen and Chen, Yulong and Wang, Longyue and Luu, Anh Tuan and Bi, Wei and Shi, Freda and Shi, Shuming},
  year={2023},
  eprint={2309.01219},
  archivePrefix={arXiv},
  primaryClass={cs.CL},
  url={https://arxiv.org/abs/2309.01219}
}

@misc{cis2024benchmarks,
  title={{CIS} Benchmarks},
  author={{Center for Internet Security}},
  year={2024},
  howpublished={\url{https://www.cisecurity.org/cis-benchmarks}},
  note={Cloud security configuration standards}
}

@misc{yang2025llmsec,
  title={{LLMs} in Software Security: A Survey of Vulnerability Detection Techniques and Insights},
  author={Sheng, Ze and Chen, Zhicheng and Gu, Shuning and Huang, Heqing and Gu, Guofei and Huang, Jeff},
  year={2025},
  eprint={2502.07049},
  archivePrefix={arXiv},
  primaryClass={cs.CR},
  url={https://arxiv.org/abs/2502.07049},
  note={Survey of LLM-based vulnerability detection}
}

@misc{zhang2025llmvuln,
  title={{LLM}-based Vulnerability Detection at Project Scale: An Empirical Study},
  author={Li, Fengjie and Jiang, Jiajun and Chen, Dongchi and Xiong, Yingfei},
  year={2026},
  eprint={2601.19239},
  archivePrefix={arXiv},
  primaryClass={cs.SE},
  url={https://arxiv.org/abs/2601.19239}
}

@misc{chen2025agenticcyber,
  title={{AgenticCyber}: A {GenAI}-Powered Multi-Agent System for Multimodal Threat Detection and Adaptive Response in Cybersecurity},
  author={Roy, Shovan},
  year={2025},
  eprint={2512.06396},
  archivePrefix={arXiv},
  primaryClass={cs.CR},
  url={https://arxiv.org/abs/2512.06396}
}

@misc{zhang2025multiagent,
  title={Open Challenges in Multi-Agent Security: Towards Secure Systems of Interacting {AI} Agents},
  author={Schroeder de Witt, Christian and Krawiecka, Klaudia and Krawczuk, Igor and Hagag, Ben and Anderson, William L. and Belcak, Peter and Bucknall, Ben and Cai, Xiaohong and Chopra, Ayush and Cohen, Doron and Del Rosario, Ron F. and Draguns, Andis and Gray, Annie and Katz, Keren and Mavroudis, Vasilios and Mink, Jaron and Motwani, Sumeet Ramesh and Petit, Jonathan and Rembeck, Leif-Sebastian and Smith, Chandler and Sotiropoulos, John and Young, Steven and Scheffler, Sarah and Llewellyn, Mary},
  year={2025},
  eprint={2505.02077},
  archivePrefix={arXiv},
  primaryClass={cs.AI},
  url={https://arxiv.org/abs/2505.02077}
}

@article{pan2024cicd,
  title={Ambush From All Sides: Understanding Security Threats in Open-Source Software {CI/CD} Pipelines},
  author={Pan, Ziyue and Shen, Wenbo and Wang, Xingkai and Yang, Yutian and Chang, Rui and Liu, Yao and Liu, Chengwei and Liu, Yang and Ren, Kui},
  journal={IEEE Transactions on Dependable and Secure Computing},
  volume={21},
  number={1},
  pages={403--418},
  year={2024},
  doi={10.1109/TDSC.2023.3253572}
}

@misc{wang2024secscore,
  title={{SecScore}: Enhancing the {CVSS} Threat Metric Group with Empirical Evidences},
  author={Santana, Miguel and Cogo, Vinicius V. and de Sá, Alan Oliveira},
  year={2024},
  eprint={2405.08539},
  archivePrefix={arXiv},
  primaryClass={cs.CR},
  url={https://arxiv.org/abs/2405.08539},
  note={Empirical CVSS validation}
}

@misc{wang2024falsep,
  title={An Empirical Study of Static Analysis Tools for Secure Code Review},
  author={Charoenwet, Wachiraphan and Thongtanunam, Patanamon and Pham, Van-Thuan and Treude, Christoph},
  year={2024},
  eprint={2407.12241},
  archivePrefix={arXiv},
  primaryClass={cs.SE},
  url={https://arxiv.org/abs/2407.12241},
  note={Accepted at ISSTA 2024}
}

@misc{anthropic2025haiku,
  author={{Anthropic}},
  title={{System Card: Claude Haiku 4.5}},
  year={2025},
  month=oct,
  howpublished={\url{https://www.anthropic.com/claude-haiku-4-5-system-card}},
  note={Accessed: 2026-10-03}
}

@misc{openai2026gpt56,
  author={{OpenAI}},
  title={{GPT-5.6 System Card}},
  year={2026},
  month=jul,
  howpublished={\url{https://deploymentsafety.openai.com/gpt-5-6}},
  note={Accessed: 2026-10-03}
}

@misc{adams2023stridegpt,
  author={Adams, Matt},
  title={{STRIDE GPT}: An {AI}-Powered Threat Modelling Tool},
  year={2023},
  howpublished={\url{https://github.com/mrwadams/stride-gpt}},
  note={MIT License. Accessed: 2026-10-03}
}

@misc{wu2024threatmodelingllm,
  title={{ThreatModeling-LLM}: Automating Threat Modeling using Large Language Models for Banking System},
  author={Wu, Tingmin and Yang, Shuiqiao and Liu, Shigang and Nguyen, David and Jang, Seung and Abuadbba, Alsharif},
  year={2024},
  eprint={2411.17058},
  archivePrefix={arXiv},
  primaryClass={cs.CR},
  url={https://arxiv.org/abs/2411.17058}
}

@article{gao2019adaptiveensemble,
  author={Gao, Xianwei and Shan, Chun and Hu, Changzhen and Niu, Zequn and Liu, Zhen},
  title={An Adaptive Ensemble Machine Learning Model for Intrusion Detection},
  journal={IEEE Access},
  volume={7},
  pages={82512--82521},
  year={2019},
  doi={10.1109/ACCESS.2019.2923640}
}

@inproceedings{wang2023selfconsistency,
  author={Wang, Xuezhi and Wei, Jason and Schuurmans, Dale and Le, Quoc and Chi, Ed and Narang, Sharan and Chowdhery, Aakanksha and Zhou, Denny},
  title={Self-Consistency Improves Chain of Thought Reasoning in Language Models},
  booktitle={International Conference on Learning Representations (ICLR)},
  year={2023}
}

@inproceedings{manakul2023selfcheckgpt,
  author={Manakul, Potsawee and Liusie, Adian and Gales, Mark J. F.},
  title={{SelfCheckGPT}: Zero-Resource Black-Box Hallucination Detection for Generative Large Language Models},
  booktitle={Proceedings of the 2023 Conference on Empirical Methods in Natural Language Processing (EMNLP)},
  year={2023}
}

@misc{du2023debate,
  author={Du, Yilun and Li, Shuang and Torralba, Antonio and Tenenbaum, Joshua B. and Mordatch, Igor},
  title={Improving Factuality and Reasoning in Language Models through Multiagent Debate},
  year={2023},
  eprint={2305.14325},
  archivePrefix={arXiv},
  howpublished={\url{https://arxiv.org/abs/2305.14325}}
}

@inproceedings{rahman2019sevensins,
  author={Rahman, Akond and Parnin, Chris and Williams, Laurie},
  title={The Seven Sins: Security Smells in Infrastructure as Code Scripts},
  booktitle={Proceedings of the 41st International Conference on Software Engineering (ICSE)},
  pages={164--175},
  year={2019},
  doi={10.1109/ICSE.2019.00033}
}

@misc{pydifflib,
  author={{Python Software Foundation}},
  title={difflib: Helpers for Computing Deltas},
  howpublished={\url{https://docs.python.org/3/library/difflib.html}},
  note={Python 3 standard library. Accessed: 2026-10-03}
}

@misc{gitea,
  author={{Gitea}},
  title={Gitea: Git with a Cup of Tea},
  howpublished={\url{https://about.gitea.com/}},
  note={Version 1.21. Accessed: 2026-10-03}
}

@misc{jenkins,
  author={{Jenkins Project}},
  title={Jenkins},
  howpublished={\url{https://www.jenkins.io/}},
  note={Version 2.440. Accessed: 2026-10-03}
}

@misc{mlflow,
  author={{MLflow Project}},
  title={{MLflow}},
  howpublished={\url{https://mlflow.org/}},
  note={Version 2.9.2. Accessed: 2026-10-03}
}

@misc{postgresql,
  author={{PostgreSQL Global Development Group}},
  title={{PostgreSQL}},
  howpublished={\url{https://www.postgresql.org/}},
  note={Version 15. Accessed: 2026-10-03}
}

@misc{clair,
  author={{Project Quay}},
  title={Clair: Vulnerability Static Analysis for Containers},
  howpublished={\url{https://github.com/quay/clair}},
  note={Version 4. Accessed: 2026-10-03}
}

@misc{redis,
  author={{Redis Ltd.}},
  title={Redis},
  howpublished={\url{https://redis.io/}},
  note={Version 7. Accessed: 2026-10-03}
}

@misc{anthropic2026sonnet5,
  author={{Anthropic}},
  title={{System Card: Claude Sonnet 5}},
  year={2026},
  howpublished={\url{https://www.anthropic.com/claude-sonnet-5-system-card}},
  note={Accessed: 2026-11-03}
}

@misc{anthropic2026opus55,
  author={{Anthropic}},
  title={{System Card: Claude Opus 5.5}},
  year={2026},
  howpublished={\url{https://www.anthropic.com/claude-opus-5-5-system-card}},
  note={Accessed: 2026-11-03}
}

@misc{anthropic2026opus48,
  author={{Anthropic}},
  title={{System Card: Claude Opus 4.8}},
  year={2026},
  howpublished={\url{https://www.anthropic.com/claude-opus-4-8-system-card}},
  note={Accessed: 2026-11-03}
}

\end{document}